# Cosmological Expansion Started from the Big Bounce on Local Rotation

Nurgaliev I.S.

Department of Physics, Moscow State Timiryazev University, Moscow, 127550, Timiryazevskaya Str., 49.

E-mail: ildus58@mail.ru

Absract

The cosmological models in the frame of the Newtonian and general relativistic treatments are considered. The centrifugal forces acting between particles rotating randomly around each other are shown to be able to reverse gravitational collapse.

Introduction

The clear and complete solution of the two-body problem hints that falling of the unlimited amount of small bodies (particles, material dots) into each other should not be common event in the given randomly distributed initial conditions as it is not for the pare of dots. The multiple scattering phenomena, or bounces without collisions, and, much rarely, emerging of the gravitationally bounded pairs and clusters should be much more common phenomena instead. But if we "recall" that real celestial bodies have the nonzero sizes, even though much less than the characteristic sizes of the existing voids, we conclude that the catastrophic collisions are inevitable, then we can propose consistent mechanism for realistic, rather than fatalistic model of the cosmologic evolution. So instead of creationistic-apocalyptic Universe we get Universe full of smaller catastrophes but realistic one. Such a dynamic simultaneously self-organizing (self-clustering) and self-destroying evolution can take place on the static (in average), as wanted Einstein previously, as well as on the expanding, as it was concluded observing luminous component, or extracting background, possibly existing invisibly so far. Local rotations (vortexes) play the role of radical stabilization of cosmological singularity in the retrospective extrapolation and making static or steady in-the average state of the universe or a local region possible. Therefore Einstein could "permit" the galaxies to rotate instead of postulating lambda-term ad hoc in the case of general relativistic consideration of static in average Universe [1]. Though, it dose not mean necessarily that the lambda-term is not needed because of other arguments.

Rehabilitation of the local rotation-vortex

Let us consider local imaginary spherical region of the homogeneous and isotropic infinite distribution of gravitating "dust". As Milne and McCrea did, we can ignore the surrounding matter thanks to Birkhoff theorem. But in contrary to Milne and McCrea [2], we do not demand the test particle rest at the contracting sphere marking the boundary of the ball of the constant mass but let it move with the typical peculiar cosmological velocity $\vec{v}_{pecular}$ on the sphere because rotation is a typical motion in the Universe along with well measured expansion after Hubble, and the galaxies do have peculiar components of their motion. In other words we rehabilitate vorticity, and thereby long time ignored centrifugal cosmological forces as well. $\vec{v}_{pecular}$ is perpendicular to pure Hubble expansion (peculiar expansion component are averaged out) and is ignored component of the cosmological motion, in particular, in the standard general relativistic Friedman-Lemaitre models as well [3]. The exclusion of this component in the standard cosmology developments takes place either by choosing synchronous commoving system of coordinates (e.g. Khalatnikov, Landay-Liphshits et. al.) or placing rotation equal to zero ad hoc (e.g. Ellis, Hawking, Penrose et. al.).

Newtonian cosmology is governed by equations with traditional notations:

$$\frac{d\vec{v}}{dt} = \frac{\partial \vec{v}}{\partial t} + \vec{v}\nabla\vec{v} = -\nabla(\varphi + \frac{1}{\rho}p) \quad (1)$$

$$\frac{\partial \rho}{\partial t} + \nabla\rho\vec{v} = 0 \quad (2)$$

$$\Delta\varphi = 4\pi G\rho \quad (3)$$

$$\frac{1}{2}grad v^2 = [\vec{v}\,rot\,\vec{v}\,] + (\vec{v}\nabla)\vec{v} \quad (4)$$

$$\frac{d\vec{v}}{dt} = \frac{\partial \vec{v}}{\partial t} + \frac{1}{2}grad v^2 - [\vec{v}rot\vec{v}] = -\nabla(\varphi + \frac{1}{\rho}p) \quad (5)$$

Based on cosmological principle and Hubble law for the averaged expansion
$\langle \vec{V} \rangle = \langle \vec{V}_{Hubble} + \vec{v}_{pecular} \rangle = H(t)\vec{R}$ we suppose that averaging procedure

$$\langle \frac{d\vec{v}}{dt} \rangle = \langle \frac{\partial \vec{v}}{\partial t} \rangle + \frac{1}{2}grad\langle v^2 \rangle - \langle [\vec{v}rot\vec{v}] \rangle = -\langle \nabla(\varphi + \frac{1}{\rho}p) \rangle \quad (6)$$

gives

$$\langle \frac{\partial \vec{v}}{\partial t} \rangle = -\nabla(\langle \varphi + \frac{1}{\rho}p + \frac{1}{2}\langle v^2 \rangle \rangle) \Rightarrow R\frac{d}{dt}H = -RH^2 - \frac{4\pi G}{3}\rho R + \frac{\langle v_p^2 \rangle}{R}$$

Substituting $\langle v_p^2 \rangle = \omega^2 R^2$, here and in the further equations omitting for compactness the averaging signs around vortex-related values − because around other values such as H and $\rho$ we traditionally omit them − we get new simple local cosmologic equation

$$\dot{H} + H^2 = \omega^2 - \frac{4\pi G}{3}\rho, \text{ where } \dot{H} = \frac{d}{dt}H. \quad (8)$$

The law of angular momentum conservation $\omega R^2 = const = K$ and of mass conservation $\frac{4\pi}{3}\rho R^3 = const = M$ makes (8) solvable:

$$\dot{H} + H^2 = K^2/R^4 - GM/R^3, \text{ or } \ddot{R} = K^2/R^3 - GM/R^2. \quad (9)$$

Another way of presentation (9) is in only local variables, beautifully:

$$\dot{H} + H^2 = K_1^2 \rho^{4/3} - (4\pi G/3)\rho,$$

Or, as an equation of the oscillator type with variable frequency

$$\ddot{R} = [R(K_1^2 \rho^{4/3} - (4\pi G/3)\rho)]R. \quad (10)$$

Here we have got good surprise. The same functional dependence of $\omega^2$ on R as of energy density and pressure of ultra-relativistic matter (electromagnetic radiation, photons gas), all of them while isotropic are proportional to $1/R^4$, and the very same law of conservation of the averaged shear squared $\sigma^2$ (the latter causes black "matter" effect) remain the functional character of (9) unchanged causing only the re-defining the constant $K^2 = \Omega^2 - \Gamma^2 - \Sigma^2$, where constants $\Omega, \Gamma, \Sigma$ stand for vortex, radiation (energy density and pressure) and shear constants in corresponding conservation laws. Verification of this is provided by Raychaudhuri equation. So, we may say that vortex acts as a negative pressure ultra-relativistic "gas" without material carrier. It is characterized by dark energy. There is nobody to radiate and, therefore, nothing to be seen. At the same time functionally radiation-like contribution to cosmological dynamics, but opposite by sign relative to the contribution of regular matter, exists in kinematical sense. It is repulsive, despite it is beautiful.

So we get the solutions differing from each other only by constants for the both Newtonian and composed of isentropic mixture of radiation and dust type matter general relativistic description ($\rho \to \rho + 3p/c^2$, where $p$ is pressure, $c$ is speed of light).

Here is the first integral of (9)
$$\frac{1}{2}H^2 = -K^2/2R^4 + GM/R^3 + \frac{A}{R^2}, \quad (11)$$
where A is a constant of integration. There are inflation-type period with slowly changing H in the beginning (after the bounce). At the earlier stages of expansion in the Universe $\omega^2$ dominates over the gravitational attraction in the regular matter and takes place acceleration at least in that period.

Let us write down final integral of the cosmologic equations:
For $A \geq 0$ we have:
$$t + t_0 = -(2A)^{-1}(2AR^2K + 2GMR - K^2)^{1/2} -$$
$$- GM(2A)^{3/2} \ln(2^{3/2} A^{1/2}(2AR^2 + 2MGR - K^2) + 4AR + 2GM),$$
For $A \leq 0$ we have
$$t + t_0 = (2A)^{-1}(2AR^2K + 2GMR - K^2)^{1/2} -$$
$$- GM(-2A)^{3/2} \arcsin[(2AR + GM)/(2A^2K^2 + GM^2)^{1/2}].$$

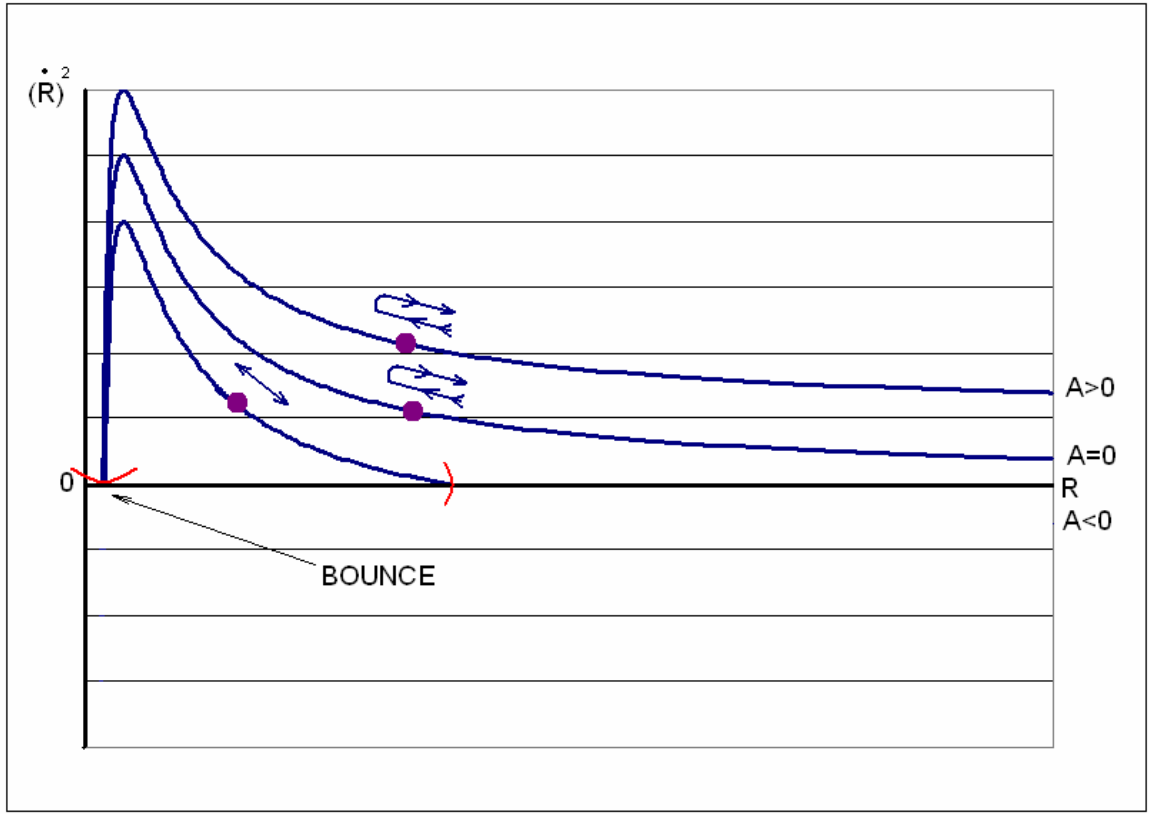

This phase diagram helps us to visualize the oscillation patterns and/or asymptotical regimes of the solutions. (Compare with Kepler's task for two bodies, Descartes' ideas of Vortex Universe and Voltaire's image of a watch that had been wound up at the creation and continues ticking — expanding-whirling — to eternity … or until wears out thermodynamically and will need in repairing and upgrading phase transition).

"How much" $\omega^2$ in amount orders is able to overcome gravitational contraction of ultra-relativistic matter (we suppose that traditional cosmology is proved science including later radiation-dominated period) guarantee the cosmologic bounce of the previously contracting

Universe before expansion? Or, what is exact value of $\omega^2$ in the radiation-dominated Einstein static Universe? We have

$$\omega^2 = \frac{4\pi G}{3}\rho_{radiation} = \frac{4\pi G}{3c^2}aT^4, \text{ where } a = \frac{\pi^2 k^4}{15\hbar^3 c^3} = 7.56\times 10^{-16} J/m^3 K^4. \text{ Therefore,}$$

$$\omega^2 = \frac{4\pi^3 G k^4}{45\hbar^3 c^5}T^4. \quad (13)$$

Or, compactly,

$$\omega^2 = a\chi T^4, \text{ or}$$
$$\omega^2 = \aleph T^4 \quad (14)$$

where $\chi = \frac{4\pi G}{3c^2} = 3.1\times 10^{-27} m/kg,$ and

$$\aleph = \frac{4\pi^3 G k^4}{45\hbar^3 c^5} = 2.11\times 10^{-41} c^{-2} K^{-4} \quad (15)$$

is a new universal constant.

Note that simple and "familiar looking" expression $\omega^2 = \frac{4\pi G}{3}\rho$ in the given new context has quite new general meaning and possess far going connotation. It is local. The illegally trampled right to rotate is returned to the point of matter. For T=2.7K we get that even such a small value as $\omega^2 \approx 1.12\times 10^{-40}$ rad$^2$/c$^2$ could be enough to compensate the radiation contribution to the cosmological contraction preceded to the observed expansion, i.e. less than, supposedly, existing value. Anyway, the candidates, such as Birch rotation and others, if reconfirmed, would be more than necessary to provide this magnitude, and to treat our Universe as oscillating around the static state, having been sort of steady state in average, until we find more fundamental concepts for thinking of Universe, or, preferably, succeed in more sophisticated observations and measurements. Specifically, in calculating $\omega^2$ from WMAP data.

Discussion.

We have given an answer "Thanks, no." to the following hypothesis. In the Universe somebody somehow with the some unknown purpose, at the some mysterious previous stage of its evolution had fine tuned with 100%-prcisness zero scattering of each particle around each another one. This Entity provided by this fantastic job the delivery all of them to the very same point at the same time. Sorry, not in this Universe. Look at the skies. They are full of rotation. So, please, let us consider $\vec{v}_{pecular}$. Even though, we do respect creationistic-apocalyptic worldview of our predecessors, This Mighty Entity, who is capable to do this fine tuning job, is hardly prone to such conspiracies intention. Because, as we know according to A. Einstein, "God is sophisticated but not malicious."